\documentclass[aps]{article}
\usepackage{amsthm}
\usepackage{epsf}
\usepackage{latexsym}
\usepackage{amsmath}
\usepackage{amsfonts}
\usepackage{amssymb}
\usepackage{graphicx}
\usepackage{graphicx,epsf}
\usepackage{dsfont}
\usepackage[]{latexsym}

\newcommand{\p}{\phantom}

\begin{document}

\title{On the equation of motion for test particles in an ambient gravitational field as the Wong equation
 for a generalized gauge theory}
 \date{}

\author{S. Fabi \thanks{Department of Physics and Astronomy, The University of Alabama, Box 870324, Tuscaloosa, AL 35487-0324, USA. email: fabi001@bama.ua.edu}~ and G.S. Karatheodoris \thanks{email: karatheodoris.george@gmail.com}}

\maketitle


\begin{abstract}
    {In this note we demonstrate the equation of motion for test particles in an ambient gravitational field for the teleparallel theory of gravity, considered as a generalized gauge theory, using a computational scheme due to Feynman.  It can be thought of as the Wong equation for a generalized gauge theory.  The Wong and Lorentz equations become identical when the generators of a generalized non-abelian gauge theory are taken to be the local translation generators.}
\end{abstract}


\section{Introduction}
The study of the mathematical and conceptual relationship between
Yang-Mills gauge theory and Einstein's general relativity has
occupied a central place in the efforts of theoretical physicists
for some time now.  From one point of view it is the problem of the
unification of forces, since three quarters of Nature is described
well by gauge theory and the other quarter is described well by the
classical theory of general relativity.  From another point
of view it is a fundamental issue in the search for a consistent
quantum theory of gravity. Gauge theoretic ideas are not only
consistent with quantum theoretic ones, but may even form an
essential part of quantum theory, for example taming the theory when
we aspire to quantize many degrees of freedom.  Even more
fundamentally, gauge systems appear immediately from quantum
dynamics in the form of Berry's connection \cite{Berry1984} and the
non-abelian connections studied by Wilczek and Zee \cite{Wilczek1984}.

The most precise relationships between gauge theory and gravity are,
in our opinion, the various relationships uncovered within the
scheme of string theory and the results on the level of the
classical actions initiated by Utiyama \cite{Utiyama1956} and
discussed by many authors. An excellent bibliography of this
literature is provided in the book \cite{Blagojevica}.  These two
approaches have one radical difference: in the former the
association between gauge and gravity observables is non-local,
whereas in the latter it has been assumed local--indeed gravity is
conceived of as a local gauge theory.  We think that this difference
is due to both the fact that the string theory approach includes
quantum corrections while the alternative approach does not, and
because, even on the classical level the approach initiated by
Utiyama is not fully developed conceptually or mathematically.  In
light of the direction that the latter approach is taking, it
seems very likely that this difference will evaporate.

In this note we calculate the equation of motion for test particles
in an ambient gravitational field, in a way that makes it clear that
it is a Wong equation for a generalized gauge theory.  We say
\emph{generalized gauge theory} because the concept of local gauge
symmetry, as it appears in Yang-Mills theory, has to be extended to
include situations in which the structure constants are functions on
space-time.  In this sense the symmetry itself is dynamically
determined--although we do not introduce the dynamics for it here.
The dynamics of the symmetry corresponds to the kinetic terms for
the gravitational field.  In order to obtain the equations of motion
for test particles in an ambient gravitational field we extend a
calculational scheme due to Feynman to incorporate
non-Abelian Yang-Mills fields.  This was started in
\cite{Tanimura1992} although some results are missing, among them
the relation between the gauge fields and field strength and the
transformation law for gauge fields.  These results and some others
are obtained before we treat the gravitational case.

The final result can be stated very simply.  The Wong equation for
Yang-Mills theory is

\begin{equation}\label{Wong2}
    \dot{I}^a - f^{ab}_{\p{ab}c}A_{b\mu}\dot{x}^\mu I^c=0.
\end{equation}
By the trivial replacements
\begin{eqnarray}\label{replacements}
    I^a &\mapsto& \mathcal{P}^a(x)  \\
    f^{ab}_{\p{ab}c} &\mapsto& T^{ab}_{\p{ab}c}(x) \\
    A_{b\mu}(x) &\mapsto& h_{b\mu}(x)
\end{eqnarray}
this equation becomes the Wong equation \emph{and} the Lorentz equation for
the teleparallel equivalent of general relativity\footnote{For a review of the teleparallel equivalent of general relativity, see \cite{Pereira2007}}, and is therefore
further equivalent to the geodesic equation.  To the right of the
arrows we have the a generalized momentum, the torsion and the
frame components respectively.  Note carefully that gravity is both a
specialization and a generalization of ordinary gauge theory:  It is
a special case due to the degeneracy of the Lorentz and Wong
equations of the relevant generalized gauge theory, and a generalization due to the space-time dependence of
what were formerly the structure constants.  Note also how that
characteristic quadratic velocity dependence of the geodesic
equation arises: one velocity already appears in the ordinary Wong
equation while the other is present because of the nature of the
generator $\mathcal{P}^a(x,p)$.

\section{Gauge Theory in Feynman's scheme} \label{YMF}
Here we review the treatment of gauge theories with gauge groups
that are the direct sum of commuting compact simple and $U(1)$ subalgebras \cite{Weinbergb}, following
\cite{Lee1990jy}, \cite{Tanimura1992} with some refinements.  As the
results are the \textbf{\emph{non-abelian analogue of the Lorentz
force law}}, the \textbf{\emph{Wong equations}} for the gauge
generators, the \textbf{\emph{Bianchi identities}} for the field
strengths, the \textbf{\emph{relation between the gauge field and
gauge field strengths}}, and the \textbf{\emph{transformation law
for the gauge fields}}, we comment here on the range of validity of
these.\footnote{The result contains a scalar field contribution to
the equations of motion as well as the usual Lorentz term.  This
term appears with its own field strength and Bianchi identity.}  This
happens because the symmetry considerations adhered to do not forbid
the appearance of fundamental scalars.  The resulting theory is only
relevant in the limit in which matter field excitations can be
treated as point particles while the Yang-Mills field is governed by
classical wave equations.  This can only happen at extremely high
energies due to the mass gap, see e.g. \cite{Witten2008} for a
review.  In simple terms, the Yang-Mills-matter system is treated in
precise analogy with the classical electrodynamics of point charges;
of course, as already stated, the physical conditions under which
this approximation is valid may be exotic.

As a starting point we consider the following fundamental
assumptions as the basis of the computational scheme employed
\footnote{$x^{\mu}(\tau)$ is a Heisenberg picture operator while,
when the $\tau$ disappears, we are discussing the Sch\"{o}dinger
picture operator.  Thus when interpreting an object like $x^\mu$ we
do not merely indicate a point on the trajectory, but a full fledged
coordinate on which a field may depend, as in $A_{\mu}(x)$.  This is
part of a deeper set of issues, falling under the rubrics
\emph{restriction to one particle operators} and \emph{stability},
which will be addressed in future work.}:

\begin{itemize}
  \item  A particle moves along a trajectory $x^{\mu}(\tau)$ in $\mathbb{R}^{3,1}$, and it possess the physical properties \textbf{mass} and \textbf{color}, represented by $m$  and $I^{a}(\tau)$ respectively. $I^{a}(\tau_{0})$ is in the Lie algebra $\mathfrak{g}$ of the gauge group $G$ for fixed $\tau_{0} \in \mathbb{R}$. \footnote{There is some confusion in \cite{Tanimura1992} surrounding the association of the parameter $\tau$ with the proper time which arises from an incorrect handling of constraints (see \cite{Green} page 19) in the Poisson bracket formalism.  This will be dealt with in connection with the stability issue alluded to above, which arises from the indefinite metric $\eta$ in the algebra.  Remarkably, Grassmann variables and supersymmetry are forced on our formalism when we try to make a Lagrangian for the Wong equations; this will be discussed in future work.}
  \item  The operators $\{ x^{\mu}(\tau), \dot{x}^{\mu}(\tau), I^{a}(\tau) \}$ satisfy the following \textbf{Yang-Mills-Feynman algebra} (which we will call $\mathfrak{I}$) of \textbf{Feynman brackets} (\textbf{F-brackets}):
      \begin{align} \label{YM_algebra}
        &[x^\mu,x^\nu] = 0 \\
        &\left[x^\mu,\dot{x}^\nu\right] = (\tfrac{-i\hbar}{m})\,\eta^{\mu\nu} \\
        &\left[I^a,I^b\right] = i \hbar f_c^{\phantom{c}ab} I^c \\
        &\left[x^\mu,I^a\right] = 0
      \end{align}
      These brackets satisfy \textbf{all the properties of standard commutators resp. Poisson brackets}, in addition to a further Leibniz rule we call the \textbf{Lie-Leibniz rule} \footnote{The implication of the assumed rule \eqref{Lib2} is significant and was not appreciated in the pioneering literature. It can be shown that a dynamical system is Hamiltonian if and only if the time derivative acts on Poisson brackets in accordance with \eqref{Lib2}, thus stopping Feynman's initial motivation in its tracks.  For a proof see e.g. \cite{Jose1998}.  This explains why  exhaustive searches for novel (non-Hamiltonian) systems within the Feynman scheme, as e.g. conducted in  \cite{Carinena1995}, turned up nothing.} (since it enforces the Liebniz rule with respect to the Lie derivative).
      \begin{equation} \label{Lib2}
        \mathcal{L}_{\partial_\tau}[A,B]=[\mathcal{L}_{\partial_\tau}A,B] + [A,\mathcal{L}_{\partial_\tau}B].
      \end{equation}

      Direct products of gauge algebras are considered later, for now we imagine $\{I^a\}$ generates a gauge algebra $\mathfrak{g}$.
    \item  The coordinate operators obey second order differential equations of motion, i.e. the Heisenberg equations of motion in the form of \textbf{Lorentz's law},
        \begin{equation} \label{SODE}
          m \ddot{x}^\mu = \mathcal{F}^\mu(x,\dot{x},I),
        \end{equation}
      where the force is assumed to be \emph{linear} in the $\{I^a\}$.
    \item  The color variables are assumed to satisfy \textbf{first order differential equations of motion}, as per usual:
        \begin{equation} \label{FODE}
          \dot{I}^a= \mathcal{H}^a(x,\dot{x},I).
        \end{equation}
      In the non-Abelian case these are called \textbf{Wong's equations}. Again, $\mathcal{H}^a(x,\dot{x},I)$ is assumed \emph{linear}.  The reason why internal variables satisfy first order equations of motion while variables associated with the base space (space-time) satisfy second order equations is simply because in the latter case the generator is the momentum, which can \emph{itself} be written as a time derivative.  Thus truly, all the equations of motion are first order equations for the relevant generators as is obvious in Hamilton's framowork. In the gravitational case, i.e. when the generator is taken to be the local translation generator and a generalized local gauge theory is constructed, the Lorentz equations and the Wong equations coincide.
\end{itemize}

Feynman's original motivation was to create a formalism robust
enough to encompass systems that are not Lagrangian or Hamiltonian.
Our motivation for adopting his approach is completely different, as
we simply wish to show that gravity's fundamental \emph{generalized
gauge origin} is rather clear in his language; similarities and
differences between the gravitational and Yang-Mills cases of gauge
theory \footnote{Here we use the term \emph{ generalized gauge
theory} in a loose way to denote a theoretical structure that will
encompass both local Yang-Mills symmetries and local translation
symmetries, which can be thought of as at the root of gravity.  We
assume little more then that the symmetry itself is a dynamical
variable through the introduction of space-time dependent structure
``constants".  It is pleasing that the generalization can be stated
in simple terms and takes Weyl's definition of local symmetry one
clear step further.} are put in stark relief.  This is just as well;
since Dyson's publication of Feynman's proof no one has succeeded in
utilizing the formalism for its original purpose because all authors
have assumed the Lie-Leibniz rule \eqref{Lib2}.   In fact one can show that \eqref{Lib2} implies the existence of an Hamiltonian. The formalism has
however been used to obtain results in the noncommutative case
\footnote{See however \cite{Carinena2006}}.

Instead of postulating the Wong equation, as in \cite{Tanimura1992},
we \emph{derive} it in complete analogy with the derivation of the
Lorentz force using (\ref{FODE}) and define the gauge potential
through
\begin{equation}\label{YMgaugepot}
    \breve{\mathcal{A}}^{\mu a}(x,\dot{x},I):= \frac{m}{i \hbar}[\dot{x}^\mu, I^a].
\end{equation}
Surprisingly, the ordinary gauge potential is obscured in
\eqref{YMgaugepot}.  The index $a$ is a matrix index of
$\mathfrak{g}$ in the adjoint representation and \emph{not} an index
specifying a Lie algebra generator.  Specifically, the usual form of
the gauge potential is related to the one in (\ref{YMgaugepot}) by
\begin{equation}\label{YMgaugepot2}
    \breve{\mathcal{A}}^{\mu a}= (\mathcal{A}^{\mu}_{\p{\mu}c})(\breve{I}^c)^a_{\p{a}b}I^b,
\end{equation}
where the over-symbol in (\ref{YMgaugepot2}) designates the
\emph{adjoint representation} of $\mathfrak{g}$, explaining its
appearance in (\ref{YMgaugepot}) as well. It is easy to see that the
Yang-Mills gauge potential does not depend on the velocity by taking
the F-bracket of (\ref{YMgaugepot}) with $\dot{x}^{\nu}$ and using
the Jacobi identity.  The linearity condition then implies
\begin{equation}\label{YMgaugepot3}
    [\dot{x}^\mu, I^a]:= \frac{i \hbar}{m}\breve{\mathcal{A}}^{\mu a}_{\p{\mu a}b}(x)I^b.
\end{equation}
Taking the F-bracket of (\ref{FODE}) with $x^{\mu}$ and using
(\ref{YMgaugepot3}) we find
\begin{equation}\label{YMgaugepot4}
    [x^\mu, \mathcal{H}^a_{\p{a}b}(x,\dot{x})]I^b=-\frac{i \hbar}{m}\breve{\mathcal{A}}^{\mu a}_{\p{\mu a}b}(x)I^b.
\end{equation}
Now we can use the linear independence of the generators and the
fact that $[x_\nu,f(\dot{x})]= -\frac{i \hbar}{m}
\partial_{\dot{x}^\nu} f(\dot{x})$; we find, after integration,
\begin{equation}\label{Wong1}
    \dot{I}^a-  \breve{\mathcal{A}}_{\mu \p{a} b}^{\p{\mu}a}(x)  \dot{x}^\mu I^b=0,
\end{equation}
which is the \emph{Wong equation}. It can be written, alternatively,
\begin{equation}\label{Wong2}
    \dot{I}^a - f_c^{\p{c}ab}A_{b\mu}\dot{x}^\mu I^c=0,
\end{equation}
using $\eqref{YMgaugepot2}$.

The \emph{field strength} corresponding to the potential just
introduced is
\begin{equation}\label{YMfieldst}
    F_{\mu\nu}(x):= -\frac{m^2}{i \hbar} [\dot{x}_\mu,\dot{x}_\nu].
\end{equation}
The components of $F_{\mu\nu}$ in the Lie algebra basis $\{I^a\}$
can be worked out in the form
\begin{eqnarray*}
  F^{\mu\nu} &=& -\frac{m^2}{i\hbar}[\dot{x}^\mu,\dot{x}^\nu] \\
   &=& \frac{m}{i \hbar} [x^{\mu}, \mathcal{F}^{\nu} ] \\
   &=& \frac{m}{i \hbar} [x^\mu, \mathcal{F}_a^\nu]I^a
\end{eqnarray*}
implying
\begin{equation}\label{fstr_compts}
    F_{\mu\nu}(x,I)=F_{a\mu\nu}(x)I^a.
\end{equation}
That this is in fact related to $\breve{\mathcal{A}}_\mu$ in the
usual way will be shown \footnote{The significance of the adjoint
representation and the peculiar form of (\ref{YMgaugepot}) were not
appreciated in \cite{Tanimura1992}, which led to difficulty in the
demonstration of this fact among others.  Of course the result is representation independent.}
shortly.  Note that the $a$ index in $\eqref{fstr_compts}$ indicates
a particular basis element of $\mathfrak{g}$ and has nothing to do
with the adjoint representation.

First we can get the \emph{Bianchi identity} for $F_{\mu\nu}$ out of
the way since it reduces to the Jacobi identity for the velocity
operators,
\begin{eqnarray}\label{Biancobi}
    [\dot{x}_\mu,[\dot{x}_\nu,\dot{x}_\rho]] + cyclic   \\
    =\frac{i \hbar}{m^2} \Big([\dot{x}_\mu,F_{\nu\rho}] + cyclic \Big) \\
    =\frac{\hbar^2}{m^3}(\mathcal{D}_\mu F_{\nu\rho} + cyclic)=0.
\end{eqnarray}
To get the last line we used the identity
$[\dot{x}_\mu,\phi_a(x)I^a]=\frac{i \hbar}{m} (\mathcal{D_\mu
\phi)}_a I^a$.

To prove that indeed the field strength introduced in
$\eqref{YMfieldst}$ corresponds to the potential introduced in
$\eqref{YMgaugepot}$ and $\eqref{YMgaugepot2}$ we can start by
taking the F-bracket of $\eqref{YMgaugepot3}$ with $\dot{x}^\nu$
\begin{equation}\label{fstrcomp1}  
    [ \dot{x}_\nu, [\dot{x}_\mu, I^a]] + \Big(\frac{i \hbar}{m} \Big)[\dot{x}_\nu, \breve{A}_{\mu \p{a} b}^{\p{\mu} a}(x)I^b]=0,
\end{equation}
and then using Jacobi on the first term and Leibniz on the second we
get
\begin{multline}\label{fstrcomp2}
    [\dot{x}_\mu, [I^a, \dot{x}_\nu]] + [I^a, [\dot{x}_\nu, \dot{x}_\mu]] + \\ \Big(\frac{i \hbar}{m} \Big) \left[ \dot{x}_\nu, \breve{A}_{\mu \p{a} b}^{\p{\mu} a}(x) \right] I^b + \Big(\frac{i \hbar}{m} \Big) \breve{A}_{\mu \p{a} b}^{\p{\mu} a} \left[ \dot{x}_\nu,I^b \right]=0
\end{multline}
leading to
\begin{multline}\label{fstrcomp3}
    \Big( \frac{-i \hbar}{m} \Big) [\dot{x}_\mu, \breve{A}_{\nu \p{a} b}^{\p{\mu} a}I^b]  -\Big(\frac{i \hbar}{m^2} \Big)[I^a,F_{\mu\nu}] + \\
    \Big(\frac{i \hbar}{m} \Big)^2 \breve{A}_{\mu,\nu \p{a} b}^{\p{\mu} a} I^b + \Big(\frac{i \hbar}{m} \Big)^2 \breve{A}_{\mu \p{a} b}^{\p{\mu} a}  \breve{A}_{\nu \p{a} c}^{\p{\mu} b} I^c=0
\end{multline}
or\footnote{Notice the factors associated with the
(anti)symmetrization of indices.}
\begin{equation}\label{fstrcomp4}
    -(\breve{F}_{\mu\nu})^a_{~c}I^c + 2(\breve{A}_{[\mu,\nu]})^a_{~b}I^b + 2\breve{A}_{(\mu ~ b}^{~~a} \breve{A}_{\nu) ~ c}^{~~b}I^c=0.
\end{equation}
Changing the dummy indices and using the linear independence of
$\{I^a\}$ we have
\begin{equation}\label{fstrcomp5}
    (\breve{F}_{\mu\nu})^a_{~b}=2(\breve{A}_{[\mu,\nu]})^a_{~b} +  2\breve{A}_{(\mu ~ c}^{~~a} \breve{A}_{\nu) ~ b}^{~~c}
\end{equation}
which is the standard relation for the field strength in terms of
the potential in Yang-Mills theory.  It was simplest to prove in the
adjoint representation but it holds for any representation; we can
simply use $\eqref{YMgaugepot2}$ and its analogue for
$(\breve{F}_{\mu\nu})^a_{~b}$ in order to convert
$\eqref{fstrcomp5}$ into the representation independent form
\begin{equation}\label{fstrcomp6}
    F_{\mu\nu}= 2A_{[\mu,\nu]} + [A_\mu,A_\nu].
\end{equation}

In order to demonstrate the \emph{Lorentz force} we introduce the
function $\mathcal{G}^{\mu}(x, \dot{x},I)=\mathcal{G}^\mu_{\p{\mu}a}
I^a$ through the following equation,
\begin{equation}\label{G}
    \mathcal{G}^{\mu}(x,I):=\mathcal{F}^\mu(x,\dot{x},I)-F^{\mu\nu}(x,I)\dot{x}_{\nu}
\end{equation}
where $\mathcal{G}^\mu_{\p{\mu}a}$ can be shown\footnote{Further
details can be found in \cite{Tanimura1992}.} to be a function of
$x$ only.  A moments reflection shows that the independence of
$\mathcal{G}$ on $\dot{x}$ is all that is needed to conclude the
Lorentz force law \footnote{Modulo the scalar field contribution
$\mathcal{G^{\mu}}(x)$.} for Yang-Mills theory:
\begin{equation}\label{YMLorentzforce}
     \mathcal{F}^\mu(x,\dot{x},I)=\mathcal{G}^{\mu}(x,I) + F^{\mu\nu}(x,I)\dot{x}_{\nu}.
\end{equation}
The very same methods used above can be used to show
\begin{equation}\label{BianchiG}
    d*\mathcal{G}(x,I)=0,
\end{equation}
the Bianchi identity for $\mathcal{G^{\mu}}(x)$, where the
derivative is the usual exterior one and the Hodge star is used.

Finally we wish to demonstrate that we are really doing Yang-Mills
gauge theory by defining the gauge transformation of the Yang-Mills
gauge potential. We define gauge transformations in a novel way and
then show that the resulting formalism and conceptual content is the
same as the one inherent in the fiber bundle interpretation of
Yang-Mills gauge theory \cite{Yang1974}.

In order to motivate the construction, consider that what we must
express in mathematical terms is the symmetry of the physics under a
certain group $G$ whose algebra is $\mathfrak{g}$.  The physics is
described by $\mathfrak{I}$ and the analytic conditions described in
the axioms, so we first have to represent the action of $G$ on
$\mathfrak{I}$; however, we can be more specific, because the
Yang-Mills gauge symmetry leaves the space-time or external
structures (sometimes called `horizontal' in the geometric language) invariant by fiat (this is the \emph{verticality}
requirement in the fiber bundle language), so we expect that that
action should only act non-trivially on $\mathfrak{g}$.  The
relevant mathematical construction should now begin to clarify, at
least insofar as it must involve the adjoint action.

To review, since every Lie algebra $\mathfrak{f}$ is a vector space
via the forgetful functor which forgets the Lie bracket, one can
consider using this vector space itself as the representation space
for a representation $\breve{\Gamma}(F)$ of the group $F$.  This can
be accomplished by applying the exponential map to the adjoint
representation $\breve{\Gamma}(\mathfrak{f})$ of $\mathfrak{f}$.
Since the Yang-Mills gauge potential, $A_{\mu}(x,I^a)$ is introduced
as a function on $\mathfrak{I}$ it will transform as well--this will
define a gauge transformation.

Let $\Lambda \in \mathfrak{g}$.  Then
\begin{equation}\label{gt1}
    \delta I^a=[\Lambda, I^a]
\end{equation}
is the variation of the basis element $I^a$ under $\Lambda$.  This
can be further expanded
\begin{align}\label{gt2}
    \delta I^a  &= [\Lambda_c I^c, I^a] \\
                &= \Lambda_c(x)[I^c,I^a]  \\
                &= \Lambda_c(x) f^{~ca}_{d} I^d  \\
                &= \breve{\Lambda}^a_{~d}(x) I^d  \\
                &=: \breve{\Lambda}^a(x).
\end{align}
Therefore,
\begin{align}\label{gt3}
    \delta\breve{A}^{\mu a} &= \frac{m}{i\hbar}~ [\dot{x}^{\mu}, \delta I^a] \\
    &= \frac{m}{i\hbar}~ [\dot{x}^\mu,\breve{\Lambda}^a].
\end{align}
On the other hand, we have to take into account the fact that
$\breve{A}^{\mu a}$ is related to $A_{\mu}(x)$ by
$\eqref{YMgaugepot2}$, and it is $\delta A_\mu(x)$ that we wish to
compare with the standard result.
\begin{align}\label{gt4}
     \delta\breve{A}^{\mu a} &= \delta((\breve{A}^{\mu})^a_{~b} I^b) \\
     &= \delta(\breve{A}^{\mu})^a_{~b} I^b + (\breve{A}^{\mu})^a_{~b} \delta I^b \\
     &= \delta(\breve{A}^{\mu})^a_{~b} I^b + (\breve{A}^{\mu})^a_{~b} \breve{\Lambda}^b
\end{align}
implies
\begin{equation}\label{gt5}
    \delta(\breve{A}^{\mu})^a_{~b} I^b = \frac{m}{i\hbar}~[\dot{x}^\mu,\breve{\Lambda}^a] - (\breve{A}^{\mu})^a_{~b} \breve{\Lambda}^b.
\end{equation}
A little computation shows that the first term on the RHS of $\eqref{gt5}$ can be written
\begin{equation}\label{gt6}
 \frac{m}{i\hbar}[\dot{x}^\mu,\breve{\Lambda}^a]= \breve{\Lambda}^a_{~b} \breve{A}^{\mu b} +
    \partial^\mu \breve{\Lambda}^a_{~b}I^b.
\end{equation}
Substituting $\eqref{gt6}$ into $\eqref{gt5}$ we get
\begin{equation}\label{gt7}
    \delta(\breve{A}^{\mu})^a_{~b} I^b =\breve{\Lambda}^a_{~b} \breve{A}^{\mu b} +
    \partial^\mu \breve{\Lambda}^a_{~b}I^b -(\breve{A}^{\mu})^a_{~b} \breve{\Lambda}^b.
\end{equation}
The first and last terms on the RHS can be combined into
\begin{equation}\label{gt8}
      [\breve{\Lambda}, \breve{A}^\mu]^a_{~b} I^b
\end{equation}
by using the definitions of $\breve{A}^{\mu b}$ and
$\breve{\Lambda}^b$. The final result is
\begin{equation}\label{gt9}
    \delta(\breve{A}^{\mu})^a_{~b}=  [\breve{\Lambda}, \breve{A}^\mu]^a_{~b} + \partial^\mu \breve{\Lambda}^a_{~b},
\end{equation}
which is in agreement with the standard gauge transformation derived
from the geometric fiber bundle approach.

Having established that the infinitesimal gauge transformation of
the Yang-Mills gauge field introduced in $\eqref{YMgaugepot}$ is
formally identical to the one introduced by Yang and Mills, one can
ask whether the equations really have the same
\emph{interpretation}.  The infinitesimal space-time translation of
a function on $\mathfrak{I}$ is generated by $\dot{x}^\mu$ and
therefore we can \emph{reinterpret} $\eqref{YMgaugepot}$ as an
infinitesimal space-time translation of the color vector $I^a$:
\begin{equation}\label{inftransI1} 
    I^a \overset{\textit{trans.}}{\longmapsto}  \breve{A}^a_{~b}(x)I^b
\end{equation}
\begin{equation}\label{inftransI2}
    [\dot{x},I^a]=[\dot{x}^\mu,I^a]\partial_\mu = \breve{A}^a=\breve{A}^a_{~b}I^b
\end{equation}
In other words the gauge potential determines the behavior of the
color vector under an infinitesimal translation in space-time--the
translated vector is simply rotated by the gauge field in the
adjoint representation.  This provides a meaning of parallelism in the geometric picture, and in this algebraic picture.

\section{Relaxing the condition of verticality and generalizing the concept of local symmetry}
In the prior section we have assumed that $\mathfrak{g}$ is the
direct sum of commuting compact simple and $U(1)$ subalgebras and
associated strictly with an `internal' space.  Here we relax those
conditions and allow the gauge algebra $\mathfrak{g}$ to act
non-trivially on the space-time variables in $\mathfrak{I}$, namely
$\{x^\mu,\dot{x}^\mu\}$.  In particular we introduce an overlap
between the ``algebra" of displacements and the gauge algebra.  We
put quotes around algebra because once we introduce the generators
of local displacements, we will see that the resulting structure is
more general than a Lie algebra, it is call a Lie algebroid.  We
will not need any deep results from Lie algebroid theory in this
work, but future work will emphasize this aspect--it is the key new
idea in our formulation of a generalized gauge theory  which
incorporates gravity.  For the purposes of this note simply imagine
the notion of local symmetry extended such that the structure
constants that determine the traditional symmetry algebra are
allowed to vary from point to point.  The varying structure
constants in the momentum algebra will turn out to be the
gravitational field.  Dynamics for this structure will be introduces in later work.

The first attempt to understand gravity in the spirit of the fiber bundle language was by Utiyama \cite{Utiyama1956}, who took the gauge algebra to be $\mathfrak{so}(3,1)$; Utiyama's intuition was that behavior associated to gravitation would arise from gauging the the algebra $\mathfrak{so}(3,1)$.  This is a natural expectation for a particle physicist since locally, where there is a negligible space-time deformation, a great deal of the structure of physical laws is governed by Lorentz invariance. Interestingly however, Utiyama had to strategically insert \emph{tetrads} in an ad hoc way in order to make the construction consistent, and we will see why.  Since that attempt it has become quite clear, starting with \cite{Kibble1961} that (at least some) behavior we think of as peculiarly gravitational is induced by local translations, not by local Lorentz rotations.  This section will provide a simple proof of this fact, not by analyzing the dynamics of the gravitational fields directly, but by considering the \emph{motion of test particles in an ambient gravitational field}, and then showing how the introduction of local translation symmetry introduces the correct equations of motion and generates the relevant gravitational potentials appearing in them.  

\subsection{Local Lorentz symmetry} \label{LLS}
Here we take $\mathfrak{g}= \mathfrak{so}(3,1)$ following
\cite{Utiyama1956} thereby obtaining for $\mathfrak{I}$
\footnote{The reason $\left[x^\mu,S^{ab}\right]=0$ is that $x^{\mu}$ are coordinates
 and do not represent the components of the position vector.  A vector character may be attributed to $x^\mu$ in flat space, but must be abandoned in the manifold model for space-time.  When an affine structure is granted to the tangent space, something called a Cartan radius vector can be introduced as a sort of local position vector--this will be introduced in later work and the issue will be revisited as to what to take for the commutator between the Cartan vector and the Lorentz generators.}
\begin{align} \label{LLFA} 
        [x^\mu,x^\nu] &= 0 \\
        \left[x^\mu,\dot{x}^\nu\right] &= \frac{-i\hbar}{m}\,\eta^{\mu\nu} \\
        \left[S^{ab},S^{cd}\right] &= f^{abcd}_{~~~~~ik} ~ S^{ik} \\
        \left[x^\mu,S^{ab}\right] &= 0
\end{align}
where the adjoint representation and the structure constants of the
Lorentz algebra are defined as
\begin{equation}\label{adjlorentz}
    (\breve{S}^{cd})^{ab}_{~~ik} := f^{abcd}_{~~~~~ik} := -i(\eta^{ac} \delta^d_{~i} \delta^b_{~k} - \eta^{bc} \delta^d_{~i} \delta^a_{~k} + \eta^{ad} \delta^c_{~i} \delta^b_{~k} - \eta^{bd} \delta^c_{~i} \delta^a_{~k}).
\end{equation}
The Lorentz potentials are introduced as
\begin{equation}\label{Lorentzpot}
    (\breve{A}^\mu)^{ab}(x,\dot{x},S) := m[x^\mu,\dot{S}^{ab}]
\end{equation}
with the same conventions as in the Yang-Mills case.  The analogue
of the Wong equations is
\begin{equation}\label{LorentzWong}
    \dot{S}^{ab}-m(\breve{A}_{\mu})^{ab}_{~~cd} \dot{x}^\mu S^{cd}=0
\end{equation}
which is the statement that the tensor $S^{ab}(\tau)$ is parallel
transported along the curve associated with the parameter of
differentiation, with respect to the Lorentz connection
$(\breve{A}_{\mu})^{ab}_{~~cd}$.  Following this path leads in the
end to a Lorentz force law of the form
\begin{equation}\label{LL} 
    \mathcal{F}^\mu(x,\dot{x},S)= \mathcal{G}^{\mu}(x,S) +F^{\mu\nu}_{~~ab}(x)\dot{x}_\nu S^{ab} ,
\end{equation}
where $\mathcal{G}$ is linear in $S^{ab}.$  The important
characteristic of $\eqref{LL}$ is its linearity in the velocity.  By
going through the derivation, it is rather difficult to imagine how
that problem is to be solved, even by the addition of the
translation generator for the Poincar\'{e} group.  The origin of
quadratic force law is quite mysterious at this point, but it is
also very necessary, in fact it has been argued \cite{Singh2001a}
that certain fundamental aspects of the gravitational \emph{field}
equations can be derived from the force law for test particles, and
that the term going as the square of the velocity is essential in
this regard. Certainly the reciprocal problem of determining the law
of motion from the field equations comprises a set of rigorous
results falling under the rubric, `\emph{the problem of motion}'. In
the problem of motion studies it is clear that the hallmark of
distinctly gravitational equations of motion is their dependence on
the \emph{square} of the velocity of the test particle.

\subsection{Direct products}
We take this opportunity to sketch the results for direct products
demonstrating that the problem will persist through their
introduction.  The purpose of this exercise is to leave one fewer in
a short list of possible solutions to the quadratic velocity problem
and to pursue it for its intrinsic interest.

Suppose we consider $\mathfrak{g}=\mathfrak{g}_1 \otimes
\mathfrak{g}_2$ with basis $\{I_i^a \otimes I_j^b\}$.  We make all the
usual assumption, e.g. linearity of all functions in the generators,
and introduce a pair of gauge potentials
$\{(\breve{A}_{\mu}^{I_i})^a_{~b}(x)\}_{i=1}^2$, where
$\{I_1^a\}=\{I^a\}$ and $\{I_2^a\}=\{J^a\}$ in the usual way and the
result is a direct product of the construction we have already
presented.  In particular, the Lorentz force law is
\begin{equation}\label{LLdirectprod}
     \mathcal{F}^\mu(x,\dot{x},I_i)= \mathcal{G}^{\mu}(x,I_i) +\sum_i F^{\mu\nu}_{i~~a}(x)\dot{x}_\nu I_i^{a}.
\end{equation}
The basic characteristics of this result would not be affected by
further products.

Thus we seem to have only a few options left
\begin{itemize}
  \item The quadratic law follows from the fact that the Poincar\'{e} group is the \emph{semi}-direct product of the translation and Lorentz groups. In other words the dynamical structure of the geodesic type equation stems from the particular \emph{way in which the translations are algebraically treated} in the Poincar\'{e} group.
  \item The field strengths and potentials are not linear in the gauge generators.  This would change the whole structure of gauge theory and might be very interesting to study.
  \item There is \emph{something unusual about the translation generators} that will somehow fix the problem.  If this is true we might consider dispensing with the Lorentz generators altogether and seeing if gravitational behavior remains---we will do this.
\end{itemize}
The third possibility is a genuine solution--the second will not be
explored here.  This was gradually understood after the Utiyama
paper. Kibble \cite{Kibble1961}, it is often said, treated
translations \emph{on a par} with Lorentz transformations by using
the entire Poincar\'{e} group.  While this is true in the sense that
he did not introduce the tetrads in an ad hoc way--they are
dynamical variables and not background structures--the significant
difference between the behavior of the tetrads and the spin
connection is not emphasized by the language ``on a par with".

Our view is that the difference is substantial and has been
underplayed.  Statements in the literature that general relativity
is a special case of gauge theory in which the fiber is the tangent
space are very common and we think incorrect.  It is an active topic
of research \cite{Tresguerres2008, Tresguerres2002} to find a proper
mathematical setting for a gauge theory involving external
(space-time) symmetries.

It should be mentioned that possibilities one and three posses some
overlap--there is no analogue of the Poincar\'{e} group in which
translations are combined with rotations via direct products,
however some peculiarly gravitational behavior still cannot be
associated with the \emph{way in which the translations are
embedded} into the Poincar\'{e} group due to the remarkable fact
that the Lorentz part of the Poincar\`{e} group can be
\emph{ignored} for the purpose of constructing the equation of
motion for spinless test particles--therefore the analysis need not be modified.

\subsection{Local translation symmetry}
The basic insight that allows the introduction of local translation
invariance, and consequently gravity into this algebraic framework
is contained in  \cite{Gronwald1995}, thus we review their argument.
It is couched in a geometric language, but it will be a simple
matter to transplant the ideas into our more general algebraic
framework.\footnote{With the advent of noncommutative geometry, and
the unification of mathematics that it entails, it is beginning to
be awkward to make such distinctions in the first place.}

Consider the space of $1$-forms with basis $\{dx^i\}$ on a manifold
$\mathcal{M}$.  The $\{dx^i\}$ are globally translation invariant,
but not locally so:
\begin{equation}\label{gloabaldx1}
    x^i \mapsto {x'}^i=x^i + \epsilon^i \quad \epsilon^i \in \mathbb{R}
\end{equation}
implies
\begin{equation}\label{globaldx2}
    dx^i \mapsto d{x'}^i=dx^i,
\end{equation}
whereas
\begin{equation}\label{localdx1}
    x^i \mapsto {x'}^i= x^i + \epsilon^i(x) \quad \epsilon^i(x) \in \mathcal{C}^{\infty}(\mathbb{R}^4)
\end{equation}
implies
\begin{equation}\label{localdx}
    dx^i \mapsto d{x'}^i= dx^i +d\epsilon^i.
\end{equation}

In this circumstance we can introduce a compensating field
$\overset{t}{A^a}(x)$ ($t$ for translations), which will allow us to
make a basis of differential forms $\{Dx^i\}$ that is invariant
under local translations, provided the potential transforms
properly,  Specifically,
\begin{equation}\label{Dx}
    \delta \overset{t}{A^a}=-\delta^a_{~i}d\epsilon^i \quad \Rightarrow \quad \delta Dx^i=0.
\end{equation}
and
\begin{equation}\label{tetrad}
    e^a=\delta^a_{~i}Dx^i=\delta^a_{~i}dx^i + \overset{t}{A^a} \quad \Rightarrow  \quad \delta  e^a=0
\end{equation}

This is simply an alternative to the usual way of introducing the
tetrad basis that emphasizes its origin as a translational
potential, in accordance with the findings of \cite{Kibble1961}.
Note that it is not quite the local translation potential, but it
\emph{is} an invariant function of it. Actually the correct
identification of the local translation potential has been a controversial issue.  The difficulty is that  the tetrad does not have the correct transformation properties, for a review see \cite{Tresguerres2002}.

In terms of the elements of $\mathfrak{I}$ we might expect the
following associations ($\looparrowright$ designates an association)
\begin{equation}\label{assoc1}
    \delta^a_{~i}dx^i \quad \looparrowright \quad \delta^a_{~\mu} \dot{x}^{\mu}
\end{equation}
and
\begin{equation}\label{assoc2}
    \delta^a_{~i}dx^i + \overset{t}{A^a} \quad \looparrowright \quad \delta^a_{~\mu} \dot{x}^{\mu} + \overset{t}{A^a}
\end{equation}
where
\begin{equation}\label{assoc3}
    \overset{t}{A^a}=\overset{t}{A^a_{~\mu}} \dot{x}^\mu
\end{equation}
on the right hand side of $\eqref{assoc2}$.  $\overset{t}{A^a}(x)$
may be thought of as some kind of gauge field fluctuation, similar
in type to the ones encountered when constructing invariant
coordinates in noncommutative theories, see e.g. \cite{Douglas2001}
p. $984$. We can now finally define a new element of $\mathfrak{I}$
\begin{equation}\label{tetradelement}
    e^a(x,\dot{x}) := \delta^a_{~\mu} \dot{x}^\mu + \overset{t}{A^a}_\mu(x)\dot{x}^\mu.
\end{equation}
Here no computation needs to be done to ensure that there is no
velocity dependence in $\overset{t}{A^a}_\mu(x)$.  It is important
to notice that the temptation to use the name $\dot{y}^a$ for $e^a$
leads to a mathematical contradiction as the fluctuation is
generally non-holonomic; there is no coordinate whose derivative
gives the right hand side of $\eqref{tetradelement}$.

We now extend the algebra $\mathfrak{I}$ by replacing $\dot{x}^\mu$ by $\eqref{tetradelement}$.
Making the definitions
\begin{align}\label{tfs2}
    -i \hbar ~ W_{\mathfrak{I}}^{ab}:={m^2}[e^a,e^b]=[\mathcal{P}^a,\mathcal{P}^b] \\
     -(\tfrac{m^2}{i \hbar})h^a_{~\mu} h^b_{~\nu}[\dot{x}^\mu,\dot{x}^\nu] = h^a_{~\mu} h^b_{~\nu}W_{\mathfrak{g}}^{\mu\nu} =: W_{\mathfrak{g}}^{ab},
\end{align}
we can determine the relationship between the various field strengths associated with the algebras $\mathfrak{g}$, $\mathfrak{I}$, and the algebra associated with the local translations $\mathfrak{b}$.\footnote{The reason we label the field strength
associated with $\mathfrak{g}$ using the letter $W$ and not $F$ is explained
in the appendix A.}
The \emph{translation field strength} or
\emph{torsion} \footnote{Sometimes the word torsion is reserved for
the translation field strength when all the indices have been
translated to Greek with the aid of the tetrad components
$h^a_{~\mu}$--by torsion we mean any form of the translation field
strength.} can be simply defined as the F-bracket of the translation covariant derivatives yielding
\begin{equation}\label{tfs3}
    \overset{t}{F^{ab}}:=  D^a \overset{t}{A^b}-D^b \overset{t}{A^a},
\end{equation}
where $D^a=h^a_{~\nu} \partial^\nu$. The $\mathfrak{g}$-field strength is the sum of any field strengths
that enter into the theory via a potential introduced in the manner
of \ref{YMF} or \ref{LLS}--we can not tell what gauge potentials it
is associated with by looking at $\eqref{tfs2}$--we also have to
know what has been taken for $\mathfrak{g}$.  Note that it is
\emph{not} associated only with internal symmetries since the field
strength for Lorentz rotations are also contained in
$F_{\mathfrak{g}}^{ab}$.  All this may be suggestive of abelian gauge theory, but this is an illusion.
In fact this is a remarkable situation in
which the usual non-abelian commutator term has exactly the same
form as the usual abelian exterior derivative term.  It is essential
to note that, since $D$ is dependent upon $\overset{t}{A}$,
the field strength is non-linear
\begin{equation}\label{tfs5}
    \overset{t}{F^{ab}}(\overset{t}{A}_1 + \overset{t}{A}_2) \neq \overset{t}{F^{ab}}(\overset{t}{A}_1) + \overset{t}{F^{ab}}(\overset{t}{A}_2).
\end{equation}
Of course this is the characteristic feature that determines the
difference in the physics of abelian and non-abelian gauge theories.
Physically it corresponds to the fact that the gauge particles carry
the charge of the gauge group, in the case of gravity this is
assured, as the alternative is the existence of an energy-less gauge
boson.  Thus from both the physical and mathematical points of view
it is quite clear that, if gravity is to be thought of as a gauge
theory, that gauge theory must have the behavior of a non-abelian
gauge theory.  We are explicit about this point due to assertions to
the contrary in the literature.


\subsection{The teleparallel equivalents of general relativity}
We attribute the contentiousness of some discussions involving the
geometric torsion to an unconscious continuation of the debate about
the new quantum mechanics.  Of course, quantum mechanics is not new,
but when it was, it was more widely thought to be capable of, or in
need of, an explanation in terms of classical concepts (see e.g.
\cite{Einstein1931,Einstein1935} \cite{Bohm1954}) than it is today
(see e.g. \cite{Hooft1996,Hooft2006}). Some physicists including
Einstein, Schr\"{o}dinger, and Weyl worked on unified field theories
with an eye to eventually replacing the explanation of the stability
of matter in terms of quantum theory by one in terms of a unified
field theory that would rely upon an ontology more familiar to the
western philosophical tradition, within which scientific progress
has achieved great heights.  These efforts at unified field theories
widely utilized the torsion, and are widely considered an
embarrassment.  This phenomena should not, of course, be confused
with differing judgements regarding the possible appearance of this
tensor in laws more fundamental than general relativity or the standard model;
neither should it be confused with differing judgements about the
future of model building with torsion in various branches of physics
from elasticity theory to quantum gravity--the latter are scientific
questions and accordingly are to be probed by experiment and
self-consistency analysis.

The usual metric formulation of general relativity has a wide number
of classically equivalent\footnote{Up to boundary terms, which
\emph{are} important but do not affect the classical equations of
motion, and thus, insofar as the motion of test particles follows
from the field equations, does not influence them either.}
formulations (and physically nontrivial extensions), one of which
has come to be called \emph{the teleparallel equivalent of general
relativity}. This latter is one type of teleparallel gravity
theory--a theory of gravity in which a notion of distant parallelism
is present \emph{ab initio}.  This is made possible due to the
vanishing of the Cartan curvature throughout space-time.  It is
necessary to make substantial conceptual shifts when passing from
general relativity to its teleparallel equivalent, but the only
shift that is immediately necessary for the understanding of the
present work is to cease thinking of gravity as a manifestation of
the curvature of space-time and think of gravity as a force again,
in consonance with the Newtonian view \cite{Knox2009}.

Another interpretation of the torsion tensor is provided in the Einstein-Cartan theory
where the spin density of matter is viewed as the source of the torsion filed \cite{DeSabbata1986}.
We agree with the criticisms pointed out by Kleinert \cite{Kleinert1998,Kleinert2010}
on the Einstein-Cartan theory and will not consider it any further.

There is however one important difference between the Newtonian
gravitational force view and the teleparallel gravitational force
view, in particular, the force of gravity in Newton's theory is not
related to the gauging of a space-time symmetry, while in the
teleparallel view it \emph{is}--it will be a long known property of
manifolds, namely nontrivial metric and torsion, as opposed to
non-trivial metric and curvature that will be considered the origin
of gravity.  Therefore, in the teleparallel equivalent one expects
there to be a gravitational Lorentz force law that governs the
motion of test particles immersed in an ambient gravitational field.
For our purposes we will not need the kinetic terms for the
gravitational field itself, but will only require the analogue of
the Lorentz force.  First we display the result \cite{Pereira2007}:
\begin{equation}\label{glorentz}
    h^a_{~\mu} \frac{du_a}{ds}= F^a_{~\mu\nu} u_a u^\nu.
\end{equation}
$F^a_{~\mu\nu}$ is the field strength for local translations, i.e.
the torsion, and of course $\eqref{glorentz}$ is quadratic in the
velocity as is appropriate for a gravitational force law.  The
$h^a_{~\mu}(x)$ are the (non-holonomic) tetrad components, that can
be broken up (non-canonically) into a holonomic and non-holonomic
part.

First we will add the \emph {generalized momenta},
\begin{equation}\label{genmom}
    \mathcal{P}^a(x,p) := m e^a(x,\dot{x}) = mh^a_{~\mu}(x) \dot{x}^\mu
\end{equation}
to the objects in $\mathfrak{I}$, of course it is only a scalar
multiple of the tetrad element written in the velocity basis. Direct
computation gives
\begin{equation}\label{xpprod}
    [x^\mu, \mathcal{P}^a(x,p)] = (-i \hbar) h^{a \mu}(x)
\end{equation}
with the consistency condition that
\begin{equation}\label{raising}
    h^{\mu a}= g^{\mu\nu}(x) h_\nu^{~a}
\end{equation}
i.e. that the metric is \emph{not} trivial.  Recall that, in the
geometric approach to teleparallel gravity, the relevant geometry is
the Weitzenb\"{o}ck one, which has nontrivial metric and torsion, as
opposed to the Riemann one, which has nontrivial metric and
curvature--thus we expect a non-trivial metric to present itself.
This arises via a deformation of the relation $[x,p] \sim \eta$.
More precisely,  we proceed by \emph{assuming} $\eqref{xpprod}$ and
$\eqref{raising}$ and \emph{calculating} $[x,p]$:
\begin{equation}\label{xpprod2}
    h_\nu^{~a} [x^\mu, p^\nu] = (- i \hbar) h^{\mu a} \quad \Rightarrow \quad [x^\mu,p^\nu]= (-i \hbar) g^{\mu\nu}(x).
\end{equation}

In this approach to gravity, it is clear that, if it is going to
work at all, it must be a very special or ``degenerate" case of
gauge theory in one way and a very significant generalization in
another.  The way in which it is special manifests itself, in our
approach, by the fact that the Wong equations for the local
translation generators $\mathcal{P}^a(x,p)$, and the Lorentz force
law for $p^\mu$ (the Heisenberg equations of motion), must
\emph{coincide}.  The notion that gravitation is a \emph{special}
case of gauge theory was considered by Yang in 1974 \cite{Yang1974, Yang1975}, among
others, and has been codified in the notion that gravity is the
gauge theory for which the fiber bundle is chosen to be the tangent
bundle.  That particular coincidence of fiber and tangent space, is
represented, in our approach, by the identity of the Wong and
Lorentz equations.

The way in which gravitation is a significant \emph{generalization}
of gauge theory is much more poorly understood, and it corresponds,
in our approach, to the fact that the structure constant are
actually \emph{functions} on the manifold
\begin{equation}\label{strfuns}
    [\mathcal{P}^a(x),\mathcal{P}^b(x)]=T^{ab}_{~~c}(x)\mathcal{P}^c(x).
\end{equation}
This can be understood as an expansion of the meaning of \emph{local
symmetry} that is \emph{different} from the one that was originally
considered by Weyl. This kind of localization of symmetry can be
considered to generalize the assumption that the symmetry of the
theory, defined by the structure constants of a Lie algebra, cannot
vary from point to point.  This is presently being examined
elsewhere and we do not need further results for the purposes of
this work.

Since the $\mathcal{P}^a(x)$ satisfy the Jacobi identity, the Bianchi identity for the torsion follows:
\begin{equation}
(D^aT^{bc}_{~~d}+T^{bc}_{~~e}T^{ae}_{~~d})+cyclic=0,
\end{equation}
where $D^a=h^{a\mu}\partial_{\mu}$.

\section{Conclusions}
We claim that when the pioneering ideas of Utiyama et. al. are
pushed forward and enriched with a new principle of local symmetry a
relatively clear theory of gauge gravity begins to emerge, the
theory appears very friendly to traditional concepts of quantization
due to its gauge flavor. We calculate the Wong equation of this
theory. The Wong equation for a gauge theory describes the time
evolution of the gauge charges associated with the particles in the
theory.  In the case of a translation gauge theory, the charges are
the `internal momenta', or `translation invariant momenta' and the
Wong equation produces an analogue to the electromagnetic Lorentz
force law for the gravitational case. This equation is familiar from
the theory of teleparallel gravity, which is a theory exactly
equivalent to Einstein's general relativity, and this equation
corresponds to the geodesic equation. There are a few obvious next
steps on this direction, one surprising feature that the newly found
Lagrangian formulation for the Wong equations, when applied to the
equation for the gauge invariant momentum, requires the introduction
of Grassmann variables.

\newpage
\appendix
\section{Proof of the gravitational analogue of the Lorentz force law} \label{proof}

Taking the time derivative of $[x^\mu, \mathcal{P}_a(x,\dot{x})] =
-i \hbar ~ h_a^{~\mu}(x)$ gives
\begin{equation}\label{GLF1} 
    [ \dot{x}^{\mu}, \mathcal{P}_a] + [x^\mu, \dot{\mathcal{P}}_a] = -i \hbar ~ \partial^{\epsilon} h_a^{~\mu} \dot{x}_{\epsilon};
\end{equation}
while multiplying this by $h^a_{~\lambda}$ yields
\begin{equation}\label{GLF2}
    h^a_{~\lambda}[\dot{x}^\mu,\mathcal{P}_a] + [x^\mu, h^a_{~\lambda}\dot{\mathcal{P}}_a] = -i \hbar ~ h^a_{~\lambda} \partial^{\epsilon} h_a^{~\mu} \dot{x}_{\epsilon}.
\end{equation}

The first term on the LHS can be written
\begin{equation}\label{GLF3}
    h^a_{~\lambda}[\dot{x}^\mu,\mathcal{P}_a]= (i\hbar)(h^a_{~\lambda}\partial^\mu h_{a\epsilon}   \dot{x}^{\epsilon} - \tfrac{1}{m} g_{\lambda\epsilon}  W_{\mathfrak{g}}^{~\mu\epsilon} )
\end{equation}

Following Feynman's logic we compute the equation of motion by
noting that
\begin{equation}\label{GLF3.5}
    [x^\mu,h^a_{~\lambda}\dot{\mathcal{P}}_a] =  -(\tfrac{i \hbar}{m})\partial_{\dot{x}}^\mu  (h^a_{~\lambda}\dot{\mathcal{P}}_a),
\end{equation}
and integrating the resulting equation with respect to the velocity.  When we do this, we will get teleparallel gravity's equation of motion for test particles in an ambient gravitational field, which is simply another form of the geodesic equation. \footnote{Keeping track of ordering yields some quantum corrections which we ignore} 

One might be tempted to suppose that the only difference between the
Yang-Mills and translation cases is the presence of $x$ dependence
on the right hand side of $\eqref{GLF2}$, however,  there is a
second subtlety, nicely treated in  \cite{Tanimura1992}, which
explains the choice of the label $W_\mathfrak{g}$, as opposed to
$F_\mathfrak{g}$, for the F-bracket of velocities.  The point is
that the F-bracket of velocities, which we call
$W_\mathfrak{g}$, is \emph{itself} velocity dependent, albeit only
linearly, whereas the combination
\begin{equation}\label{GLF4}
    F_{\mathfrak{g}}^{\mu\nu}(x):= W_{\mathfrak{g}}^{\mu\nu}(x,\dot{x}) - m(\partial^\nu g^{\lambda\mu} - \partial^\mu g^{\lambda\nu})\dot{x}_\lambda
\end{equation}
is independent of the velocity.  Naturally we reserve the label
$F_{\mathfrak{g}}$ for the (velocity independent) field strength
associated with the algebra $\mathfrak{g}$.

After expanding the derivatives of the metric, we rewrite this using
the Weitzenb\"{o}ck connection
\begin{equation}\label{GLF8.5}
    \Gamma^\lambda_{~\mu\nu}:=h_a^{~\lambda} \partial_\nu h^a_{~\mu},
\end{equation}

\begin{align}\label{GLF8.6}
    W_{\mathfrak{g}}^{\lambda\sigma} = F_{\mathfrak{g}}^{\lambda\sigma} &-m g^{\sigma\delta} g^{\epsilon\eta} \Gamma^\lambda_{\p{\lambda}\eta\delta} \dot{x}_{\epsilon} -m g^{\sigma\delta} g^{\lambda\eta} \Gamma^\epsilon_{\p{\epsilon}\eta\delta} \dot{x}_{\epsilon} \\ \nonumber
    &+ m g^{\lambda\delta} g^{\epsilon\eta} \Gamma^\sigma_{\p{\sigma}\eta\delta} \dot{x}_{\epsilon} + m g^{\lambda\delta} g^{\sigma\eta} \Gamma^\epsilon_{\p{\epsilon}\eta\delta} \dot{x}_{\epsilon}.
\end{align}

The remaining $h\partial h$ terms in  \eqref{GLF2} and \eqref{GLF3}
can be written in terms of connections as
\begin{gather}\label{GLF13.5}
    h^a_{~\lambda} \partial^{\epsilon} h_a^{~\mu} = -g^{\epsilon\sigma} \Gamma^\mu_{\p{\mu} \lambda\sigma}\\
    h^a_{~\lambda}\partial^\mu h_{a\epsilon} = g_{\lambda\eta} g^{\mu\delta} \Gamma^\eta_{\p{\eta} \epsilon\delta}.
\end{gather}

Using this along with \eqref{GLF8.5}, we find the relation
\begin{align}\label{GLF14}
    -(\tfrac{i \hbar}{m})\partial_{\dot{x}}^\mu  (h^a_{~\lambda}\dot{\mathcal{P}}_a) = ~& i\hbar~ g^{\epsilon\sigma} \Gamma^\mu_{\p{\mu} \lambda\sigma} \dot{x}_{\epsilon}\\ \nonumber
    & - i\hbar~ g_{\lambda\eta} g^{\mu\delta} \Gamma^\eta_{\p{\eta} \epsilon\delta} \dot{x}^\epsilon\\ \nonumber
    &+(\tfrac{i\hbar}{m})~g_{\lambda\epsilon} \big\{ F_{\mathfrak{g}}^{\mu\epsilon}\\ \nonumber
    &-m g^{\epsilon\delta} g^{\rho\eta} \Gamma^\mu_{\p{\mu}\eta\delta} \dot{x}_\rho\\ \nonumber
    &-m g^{\epsilon\delta} g^{\mu\eta} \Gamma^\rho_{\p{\rho}\eta\delta} \dot{x}_\rho\\ \nonumber
    &+m g^{\mu\delta} g^{\rho\eta} \Gamma^\epsilon_{\p{\epsilon}\eta\delta} \dot{x}_\rho\\ \nonumber
    &+m g^{\mu\delta} g^{\epsilon\eta} \Gamma^\rho_{\p{\rho}\eta\delta} \dot{x}_\rho \big\}. \nonumber
\end{align}

The second and sixth term on the RHS cancel identically, while using the definition
\begin{equation}\label{GLF15}
 T^{\mu}_{~\nu\rho}:= \Gamma^{\mu}_{~\rho\nu}-\Gamma^{\mu}_{~\nu\rho}
\end{equation}
the first with fourth terms, and the second with the last term combines respectively to give two torsion terms and \eqref{GLF14} becomes:
\begin{equation}\label{GLF16}
-\tfrac{1}{m}\partial_{\dot{x}}^\mu  (h^a_{~\lambda}\dot{\mathcal{P}}_a) = g_{\lambda\epsilon}F_{\mathfrak{g}}^{\mu\epsilon}
-g^{\rho\eta}T^{\mu}_{~\lambda\eta}\dot{x}_{\rho}-g^{\mu\eta}T^{\rho}_{~\lambda\eta}\dot{x}_{\rho}.
\end{equation}

The integration of \eqref{GLF16} respect to $\dot{x}_\mu$ gives
\begin{equation}\label{GLF17}
\tfrac{1}{m}h^a_{~\lambda}\dot{P}_a =
T^{\epsilon}_{~\lambda\delta}\dot{x}_{\epsilon}\dot{x}^{\delta}-F_{\mathfrak{g}\delta\lambda}\dot{x}^{\delta},
\end{equation}
and being the translation field strength:
\begin{equation}\label{GLF18}
    \overset{t}{F^a_{~\mu\nu}} = h^a_{~\lambda} T^{\lambda}_{~\mu\nu}
\end{equation}
it follows that \eqref{GLF17} is in accordance with \eqref{glorentz}.


\bibliographystyle{plain}
\bibliography{GaugeGravity}

\end{document}